\documentclass[11pt]{article}
\usepackage[utf8]{inputenc}
\usepackage[T1]{fontenc}
\usepackage{lmodern}
\usepackage{textcomp}
\usepackage[margin=1in]{geometry}
\usepackage{microtype}
\usepackage{amsmath,amssymb}
\usepackage{enumitem}
\usepackage{graphicx}
\usepackage{float}
\usepackage{booktabs,tabularx,array}
\newcolumntype{L}[1]{>{\raggedright\arraybackslash}p{#1}}
\usepackage{tikz}
\usetikzlibrary{calc,arrows.meta,positioning,shapes.misc,fit}
\tikzset{
  lane/.style={draw, rounded corners, inner sep=6pt, fill=black!3},
  flowstep/.style={draw, rounded corners, align=center, inner sep=6pt, fill=white},
  acct/.style={draw, rounded corners, align=left, inner sep=4pt, fill=black!5},
  flowarrow/.style={-Latex, thick}
}
\usepackage[numbers,sort&compress]{natbib}
\usepackage[hidelinks]{hyperref}
\newcounter{definition}
\newcounter{lemma}
\newcounter{theorem}
\newcounter{invariant}
\newenvironment{definition}[1][]%
  {\refstepcounter{definition}\par\noindent\textit{Definition~\thedefinition. #1}\ }{\par}
\newenvironment{lemmaenv}[1][]%
  {\refstepcounter{lemma}\par\noindent\textit{Lemma~\thelemma. #1}\ }{\par}
\newenvironment{theoremenv}[1][]%
  {\refstepcounter{theorem}\par\noindent\textit{Theorem~\thetheorem. #1}\ }{\par}
\newenvironment{invarenv}[1][]%
  {\refstepcounter{invariant}\par\noindent\textit{Invariant~\theinvariant. #1}\ }{\par}

\newcommand{\Fr}{\mathbb{F}_r}
\newcommand{\Keccak}{\mathrm{Keccak}\text{-}256}
\newcommand{\tofield}{\mathrm{toField}}
\newcommand{\Hfn}{\mathcal{H}}

\title{\textbf{ZK Coprocessor Bridge: Replay-Safe Private Execution\\ from Solana to Aztec via Wormhole}}
\author{Jotaro Yano \\ \small Independent Researcher, Japan \\ \href{mailto:jotaro.yano@jotaro-yano.org}{\texttt{jotaro.yano@jotaro-yano.org}}}
\date{October 21, 2025}

\begin{document}
\maketitle

\begin{abstract}
\noindent
We formalize a cross-domain ``ZK coprocessor bridge'' that lets Solana programs request private execution on Aztec L2 (via Ethereum) using Wormhole Verifiable Action Approvals (VAAs) as authenticated transport. The system comprises: (i) a Solana program that posts messages to Wormhole Core with explicit finality, (ii) an EVM \emph{Portal} that verifies VAAs and enforces a replay lock, and that exposes a normative interface which parses a bound payload \texttt{secretHash}$\parallel m$ from the attested VAA, derives a \emph{domain-separated} field commitment, and enqueues an L1$\to$L2 message into the Aztec Inbox (our reference implementation (v0.1.0) currently uses \texttt{consumeWithSecret(vaa, secretHash)}; we provide migration guidance to the payload-bound interface), (iii) a minimal Aztec contract that consumes the message privately, and (iv) an off-chain relayer that ferries VAAs and can record receipts on Solana. We present state machines, message formats, and proof sketches for replay-safety, origin authenticity, finality alignment, parameter binding (no relayer front-running of Aztec parameters), privacy, idempotence, and liveness. Finally, we provide a concise \emph{Reproducibility} note with pinned versions and artifacts to replicate a public testnet run.
\par\vspace{0.6em}
\end{abstract}

\newpage

\section{Introduction}
\label{sec:intro}

Solana offers high-throughput, low-latency execution \cite{yakovenko_solana_wp} and a mature developer stack (e.g., Anchor \cite{anchor_lang_crate}, core documentation \cite{solana_docs}). Many applications, however, require \emph{private} or \emph{selectively disclosed} computation that is impractical to run directly on Solana consensus. In parallel, Aztec provides a privacy-preserving L2 with succinct rollup proofs \cite{aztec_docs}, while Wormhole offers cross-chain attestation via VAAs \cite{wormhole_docs}. This paper specifies a minimal Solana$\to$Aztec ``ZK coprocessor bridge'' that composes these systems.

\paragraph{Contributions.} (1) A normative interface where the Portal accepts a VAA and parses \texttt{secretHash}$\parallel m$ from the attested payload, deriving a domain-separated field commitment $c$ on-chain; (2) small state machines with invariants for replay-safety and single-consumption; (3) proof sketches for parameter binding, authenticity, finality alignment, privacy, idempotence, and liveness; (4) engineering guidance and a reproducibility note derived from the open-source repository \cite{zk-coprocessor-bridge}.

\paragraph{Positioning vs.\ related systems.} Our bridge externalizes private compute to Aztec and returns receipts; it does not verify heavy ZK proofs on Solana L1. This is complementary to Solana-native ZK efforts: Light Protocol's zk-compression targets state compression and privacy within Solana's data model \cite{light_protocol_whitepaper}; Arcium focuses on off-chain confidentiality with Solana integrations \cite{arcium_docs}. On the Ethereum side, Succinct provides general-purpose coprocessors and on-chain proof verification \cite{succinct_docs}; RISC Zero exposes a zkVM for arbitrary guest code with proof generation and verification \cite{risc0_docs}; Axiom offers proofs over Ethereum state and custom logic \cite{axiom_docs}. Our design is narrower: message-authenticated private dispatch from Solana to Aztec with explicit parameter binding and replay locks. Cryptographic underpinnings (e.g., PLONK/Groth16, FRI families) are orthogonal to the bridge; we reference them for completeness \cite{plonk_eprint_2019}. See also \cite{groth_2016}. See \cite{fri_iopp_2018}. See \cite{deep_fri_2019}.

\section{Model and Serialization Conventions}
\label{sec:model}

\paragraph{Domains and principals.}
There are four domains:
(D1) \textbf{Solana} origin program $P$ that posts Wormhole messages;
(D2) \textbf{Wormhole} guardians that sign VAAs;
(D3) \textbf{Ethereum} \emph{Portal} contract and the Aztec \emph{Inbox};
(D4) \textbf{Aztec L2} consumer contract $C$ that consumes L1$\to$L2 messages privately.
A relayer $R$ is any off-chain agent that submits VAAs to the Portal and may drive Aztec consumption.

\paragraph{Adversary.}
A rushing network adversary can front-run and reorder public transactions on all chains and attempt replays.
Guardian keys and the configured Solana emitter are uncompromised.
At least one honest relayer (or user) eventually submits available VAAs.

\paragraph{Cryptographic assumptions.}
$\Keccak$ is collision- and (second-)preimage-resistant.
$\Hfn$ is preimage-resistant (used to derive a secret hash).
$\Fr$ is the BN254 scalar field with prime modulus $r$.
$\tofield:\{0,1\}^{256}\to\Fr$ is canonical modular reduction of 256-bit strings into $\Fr$.
A VAA is $\mathrm{VAA}=(\textsf{body},\textsf{sigs})$, where \textsf{body} includes \textsf{emitterChainId}, \textsf{emitterAddress}, \textsf{sequence}, and \textsf{payload}.

\paragraph{Domain separation (recommended).}
Define
\[
  \mathsf{dom} := \Keccak\!\big(\texttt{``ZKCB/v1''} \parallel \textsf{emitterChainId} \parallel \textsf{emitterAddress} \parallel \textsf{sequence}\big),
\]
and the field commitment for user payload $m$ as
\[
  c := \tofield\!\big(\Keccak(\mathsf{dom}\parallel m)\big).
\]
This avoids cross-context ambiguity and reduces low-entropy risks for $m$.

\paragraph{Serialization convention (normative).}
Concatenation is $\parallel$.
Hashes output 32~bytes; all 32-byte strings are interpreted as \emph{big-endian} integers before $\tofield$.
All fixed-size integers are encoded as \emph{big-endian} when hashed by $\Keccak$.
Where explicit serialization is needed inside payloads, we use little-endian for numeric fields only where stated.

\section{Interfaces, Message Formats, and State Machines}
\label{sec:ifaces}

\subsection{Solana $\to$ Wormhole interface}
The Solana program exposes:
\begin{itemize}[leftmargin=1.25em]
  \item \texttt{post\_wormhole\_message(batch\_id, payload, finality\_flag)}:
  signs with a fixed emitter PDA and CPI-calls Wormhole \texttt{post\_message}, selecting \textsf{Confirmed} or \textsf{Finalized} finality \cite{wormhole_docs}. See \cite{solana_docs}.
\end{itemize}
The emitted VAA body includes \textsf{emitterChainId}, \textsf{emitterAddress} (the emitter PDA as 32~bytes), \textsf{sequence} (monotonic per emitter), and \textsf{payload}.

\subsection{Bound VAA payload (normative)}
\textit{This section defines the normative interface. The repository at v0.1.0 uses} \texttt{consumeWithSecret(vaa, secretHash)}\textit{; } \texttt{consume(bytes)} \textit{in v0.1.0 is a compatibility helper with }$\texttt{secretHash}=0$\textit{. New deployments SHOULD adopt the payload-bound } \texttt{consume(bytes)} \textit{that parses }$\texttt{secretHash}\parallel m$\textit{.}

For parameter binding, standardize the VAA payload as
\[
  \textsf{payload} := \texttt{secretHash} \parallel m,
\]
where $\texttt{secretHash} = \Hfn(s)$ for a secret $s\in\Fr$ and $m$ is the user message (opaque to the Portal).
\emph{Normative API.} The Portal's default entry point is
\texttt{consume(bytes encodedVaa)} that parses \texttt{secretHash} from the signed payload and computes $c=\tofield(\Keccak(\mathsf{dom}\parallel m))$ on-chain.
The legacy \texttt{consumeWithSecret(vaa, secretHash)} is retained only for backward compatibility and should be avoided in new deployments.

\subsection{Portal (EVM) API and events}
The Portal exposes:
\begin{itemize}[leftmargin=1.25em]
  \item \texttt{consume(bytes encodedVaa)}:
  verifies the VAA, checks the configured origin, enforces the replay lock on $h:=\mathrm{VAA}.\textsf{hash}$, parses $\texttt{secretHash}\parallel m$, computes $c$, and enqueues $(c,\ \texttt{secretHash})$ to the Aztec Inbox for the configured L2 instance and rollup version \cite{aztec_docs}.
\end{itemize}
It emits concise events:
\begin{itemize}[leftmargin=1.5em]
  \item \emph{VaaConsumed}$(h,\ \text{seq},\ \text{payload},\ \text{aztecKey})$,
  \item \emph{InboxEnqueued}$(h,\ \text{seq},\ c,\ \text{key},\ \text{leafIndex},\ \text{secretHash})$.
\end{itemize}

\subsection{Aztec (L2) consumer}
The Aztec contract $C$ stores the Portal's L1 address (via \nolinkurl{set_portal_once}) and exposes \nolinkurl{consume_from_inbox(content_hash, leaf_index, secret)} as the entry point.
It then calls \nolinkurl{consume_l1_to_l2_message(content_hash, secret, sender=portal, leaf_index)} and records the last observed values (content, leaf index, secret, and a counter).

\subsection{Optional receipt path (normative encoding)}
After Aztec inclusion/consumption, the Portal may publish a receipt VAA carrying:
\begin{itemize}[leftmargin=1.25em]
  \item \textsf{origEmitterChain}, \textsf{origEmitter}, \textsf{origSequence};
  \item $c$, \textsf{aztecKey}, \textsf{leafIndex};
  \item \textsf{secretHash}, \textsf{resultHash}.
\end{itemize}

\noindent\emph{Canonical encoding (normative).}
\texttt{payload} :=
\texttt{version} (uint8) $\parallel$
\texttt{origEmitterChain} (uint16 BE) $\parallel$
\texttt{origEmitter} (bytes32) $\parallel$
\texttt{origSequence} (uint64 BE) $\parallel$
$c$ (bytes32) $\parallel$
\texttt{aztecKey} (bytes32) $\parallel$
\texttt{leafIndex} (uint256 BE) $\parallel$
\texttt{secretHash} (bytes32) $\parallel$
\texttt{resultHash} (bytes32).

\medskip
\noindent\emph{Solana-side recording.}
On Solana, the recorder function \nolinkurl{record_receipt_from_vaa} should (and in v0.1.0 does not yet) require:
(i) PostedVAA account owner is Wormhole Core;
(ii) \texttt{origEmitterChain} and \texttt{origEmitter} match the local allowlist;
(iii) \texttt{origSequence} matches the PDA key;
(iv) \texttt{version}=1.

\subsection{State machines}
\textit{Portal state.}
$\sigma_{\textsf{P}}=(\textsf{consumed},\ \textsf{emitterChain},\ \textsf{emitter},\ \textsf{inbox},\ \textsf{l2Instance},\ \textsf{rollupVersion})$, where \textsf{consumed} is a set of VAA body hashes.

\textit{Transition \textsf{Consume}$(\mathrm{VAA})$.}
Preconditions:
\begin{enumerate}[leftmargin=1.25em]
  \item $\textsf{Verify}(\mathrm{VAA})=\textsf{true}$ (signatures valid; body well-formed).
  \item $\mathrm{VAA}.\textsf{emitterChainId}=\textsf{emitterChain}$ and $\mathrm{VAA}.\textsf{emitterAddress}=\textsf{emitter}$.
  \item $h:=\mathrm{VAA}.\textsf{hash}\notin\textsf{consumed}$.
\end{enumerate}
Actions:
\begin{enumerate}[leftmargin=1.25em]
  \item $\textsf{consumed}\leftarrow \textsf{consumed}\cup\{h\}$.
  \item Parse $\textsf{payload}=\texttt{secretHash}\parallel m$.
  \item Compute $\mathsf{dom}$ and $c=\tofield(\Keccak(\mathsf{dom}\parallel m))$.
  \item Enqueue $(c,\ \texttt{secretHash})$ into \textsf{inbox}.
  \item Emit events.
\end{enumerate}

\begin{invarenv}[Portal replay lock]
\label{inv:replay}
For all $h$, if $h\in\textsf{consumed}$ then any subsequent \textsf{Consume} with the same $h$ is rejected atomically.
\end{invarenv}

\textit{Aztec consumer state.} $\sigma_{\textsf{A}}=(\textsf{portalAddr},\ \textsf{lastContent},\ \textsf{lastLeaf},\ \textsf{lastSecret},\ \textsf{count})$.

\textit{Transition \textsf{ConsumeInbox}$(c,\ \ell,\ s)$.}
Preconditions: the L1$\to$L2 message $(c,\ s,\ \textsf{sender}=\textsf{portalAddr},\ \ell)$ exists and is unconsumed.
Actions: consume the message; set \textsf{lastContent}$\leftarrow c$, \textsf{lastLeaf}$\leftarrow \ell$, \textsf{lastSecret}$\leftarrow s$, and \textsf{count}$\leftarrow \textsf{count}+1$.

\begin{invarenv}[Aztec single consumption]
\label{inv:single}
Each L1$\to$L2 message can be consumed at most once.
\end{invarenv}

\subsection{Overview diagram}
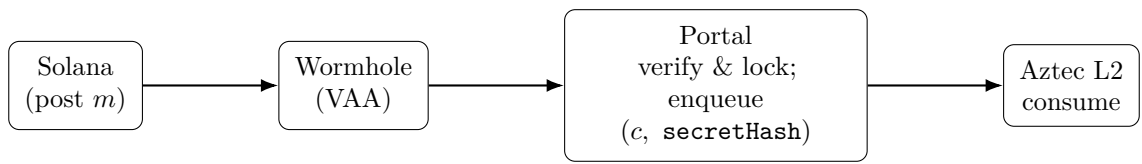
\begin{figure}[H]
\centering
\begin{tikzpicture}[font=\small, node distance=18mm]
  \node (s) [flowstep] {Solana\\(post $m$)};
  \node (w) [flowstep, right=of s] {Wormhole\\(VAA)};
  \node (p) [flowstep, right=of w, text width=36mm] {Portal\\verify \& lock;\\enqueue $(c,\ \texttt{secretHash})$};
  \node (a) [flowstep, right=of p] {Aztec L2\\consume};

  \draw[flowarrow] (s) -- (w);
  \draw[flowarrow] (w) -- (p);
  \draw[flowarrow] (p) -- (a);
\end{tikzpicture}
\caption{Minimal flow: Solana emits a Wormhole VAA; the Portal verifies and enqueues; Aztec consumes privately.}
\end{figure}

\section{Security Properties}
\noindent The results in this section assume the \emph{normative binding interface}: the Portal extracts \texttt{secretHash} from the signed payload and computes $c=\tofield(\Keccak(\mathsf{dom}\parallel m))$ with domain separation as in Section~\ref{sec:model}. When using the \emph{legacy} \texttt{consumeWithSecret(vaa, secretHash)} (as in repository v0.1.0), the \emph{parameter-binding and anti-front-running} guarantees do not hold as stated; see Section~\ref{sec:threats}.

\label{sec:security}

\subsection{Parameter binding}
\begin{definition}[Parameter binding]
A Portal implementation is parameter-binding if, for any accepted VAA with hash $h$, the pair $(c,\ \texttt{secretHash})$ enqueued to Aztec is a deterministic function of the VAA body and cannot be changed by a relayer.
\end{definition}

\begin{lemmaenv}[Binding via payload extraction]
\label{lem:binding}
If the Portal extracts \texttt{secretHash} from the VAA payload and computes $c=\tofield(\Keccak(\mathsf{dom}\parallel m))$ from the same payload, then parameter binding holds.
\end{lemmaenv}

\begin{theoremenv}[No relayer front-running of Aztec parameters]
\label{thm:nofront}
Under the model in Section~\ref{sec:model} and the state machines in Section~\ref{sec:ifaces}, a relayer cannot cause the Portal to accept a VAA with hash $h$ and enqueue $(c',\ \texttt{secretHash}')\neq(c,\ \texttt{secretHash})$ for that $h$.
\end{theoremenv}

\subsection{Replay-safety and authenticity}
\begin{definition}[Replay]
Submitting the same VAA body (same $h$) more than once to obtain multiple enqueues.
\end{definition}
\begin{theoremenv}[Replay-safety at the Portal]
\label{thm:replay}
Invariant~\ref{inv:replay} implies replay-safety: once $h$ is in \textsf{consumed}, any second \textsf{Consume} with the same $h$ is rejected.
\end{theoremenv}

\begin{definition}[Origin authenticity]
All accepted VAAs must have the configured \textsf{emitterChainId} and \textsf{emitterAddress}.
\end{definition}
\begin{theoremenv}[Authenticity]
\label{thm:auth}
Since verification succeeds only for correctly signed VAAs and the Portal enforces the origin pair, all accepted VAAs are authentic to the designated Solana program.
\end{theoremenv}

\subsection{Finality, privacy, and idempotence}
\begin{theoremenv}[Finality alignment]
\label{thm:finality}
If \textsf{Verify} only succeeds for VAAs produced under Wormhole's stated finality rules, then Portal acceptance is aligned with the origin's finality.
\end{theoremenv}

\begin{definition}[Secret privacy]
Given $\texttt{secretHash}=\Hfn(s)$, it is infeasible to recover $s$ from public data.
\end{definition}
\begin{lemmaenv}
\label{lem:privacy}
Under preimage resistance of $\Hfn$, secret privacy holds.
\end{lemmaenv}

\begin{definition}[Commitment integrity]
It is infeasible to find $(\mathsf{dom},m)\neq(\mathsf{dom}',m')$ such that $\tofield(\Keccak(\mathsf{dom}\parallel m))=\tofield(\Keccak(\mathsf{dom}'\parallel m'))$.
\end{definition}
\begin{lemmaenv}
\label{lem:binding2}
Under collision resistance of $\Keccak$ and canonical reduction to $\Fr$, collisions are negligible (up to birthday bounds).
Fixing $\mathsf{dom}$ per origin further reduces cross-context risk.
\end{lemmaenv}

\begin{theoremenv}[Privacy \& binding on Aztec]
\label{thm:privacy}
Aztec only observes $(c,\ \texttt{secretHash})$.
Under Lemmas~\ref{lem:privacy} and \ref{lem:binding2}, $c$ is binding to $(\mathsf{dom},m)$ and $s$ remains hidden given $\texttt{secretHash}$.
\end{theoremenv}

\begin{definition}[Idempotence]
For a given Solana sequence, there is at most one enqueue on Aztec and at most one L2 consumption.
\end{definition}
\begin{theoremenv}
\label{thm:idempotence}
Idempotence follows from Portal replay-safety (Theorem~\ref{thm:replay}) and Aztec single-consumption (Invariant~\ref{inv:single}).
\end{theoremenv}

\section{Liveness, Costs, and Operational Considerations}
\label{sec:liveness-costs}

\paragraph{Liveness and fairness.}
There exists at least one honest relayer (or user) that eventually submits each available VAA to the Portal; Aztec eventually includes L1$\to$L2 messages; and transactions are eventually mined on EVM and sequenced on Aztec. Under these conditions, \textsf{Consume} eventually fires; if the Aztec Inbox includes the message and the consumer submits \textsf{ConsumeInbox}, private consumption eventually occurs. Concurrent relayers may race; at most one succeeds per $h$ by Invariant~\ref{inv:replay}.

\paragraph{Portal side (EVM) costs.}
The \textsf{Consume} call cost is driven by: Wormhole verification (approximately linear in guardian signatures), event emission (linear in payload size), and the Aztec Inbox call (constant for a fixed ABI). Since $c$ derives from a single $\Keccak$ over $(\mathsf{dom}\parallel m)$, hashing cost is linear in $|m|$; in practice, $m$ is typically small.

\paragraph{Relayer and Aztec.}
Relayer latency hinges on VAA availability and EVM inclusion.
A retry loop is necessary for Aztec consumption because membership proofs become valid only after the Inbox leaf is included in a rollup block; bounded exponential backoff suffices under network liveness.
Aztec-side cost is dominated by L2 proof verification and any state writes in the consumer.

\section{Implementation Mapping and Migration Guidance}
\label{sec:mapping}

\begin{table}[H]
\centering
\small
\caption{Mapping from formal elements to implementation hooks and required deltas.}
\label{tab:mapping}
\begin{tabularx}{\linewidth}{L{0.42\linewidth} L{0.54\linewidth}}
\toprule
\textbf{Formal element} & \textbf{Implementation note} \\
\midrule
Solana message posting & Program posts Wormhole messages with explicit finality; emitter PDA fixed. \\
Portal verification \& origin & Parse-and-verify VAA; require configured emitter chain/address. \\
Replay lock & Maintain consumed-hash set; reject duplicates atomically. \\
Parameter binding (delta) & Parse \texttt{secretHash}$\parallel m$; compute $c$ from domain-separated hash; deprecate external \texttt{secretHash}. \\
Aztec enqueue & Send $(c,\ \texttt{secretHash})$ to Aztec Inbox for configured instance/version. \\
Receipts (optional) & Publish receipt VAA; on Solana, record via \nolinkurl{record_receipt_from_vaa}. \\
\bottomrule
\end{tabularx}
\end{table}

\paragraph{Domain tags and encodings.}
Adopt a fixed domain tag (\texttt{``ZKCB/v1''}) and include \emph{chain}, \emph{emitter}, and \emph{sequence}.
Serialize \textsf{payload} as \texttt{secretHash||m} with \texttt{secretHash} 32~bytes and $m$ application-defined.

\paragraph{Bound-parameter migration.}
Switch callers to \texttt{consume(bytes)}; parse \texttt{secretHash} internally; compute $c$ from $(\mathsf{dom}\parallel m)$; keep legacy path behind a feature flag for transition.

\paragraph{Receipts.}
If enabled, require the origin triple and version to match before recording.

\section{Reproducibility}
\label{sec:repro}

This note summarizes how to replicate a public-testnet run of the bridge.

\medskip
\noindent\textit{Pinned stack.}
Node 24+, Foundry, Anchor v0.31.1, Solana CLI, Aztec CLI and \texttt{aztec-wallet}, Ethers v5, Wormhole SDK v0.10.18, Aztec packages v2.0.2. Networks: Solana devnet, Wormhole testnet, Ethereum Sepolia, Aztec testnet.

\medskip
\noindent\textit{High-level steps.}
(1) Deploy the Solana program that posts Wormhole messages and note the emitter PDA.
(2) Deploy the Portal on Sepolia with the Wormhole core, emitter, and Aztec endpoints configured.
(3) Deploy the Aztec consumer and call \nolinkurl{set_portal_once}.
(4) Post a message on Solana; run a relayer to submit the VAA to the Portal and trigger Aztec consumption.
(5) Optionally publish and record a receipt VAA on Solana.

\medskip
\noindent\textit{Artifacts.}
Repository, logs, and proof pointers are in \cite{zk-coprocessor-bridge}. Observe Portal events (\emph{VaaConsumed}, \emph{InboxEnqueued}) and an incremented counter on the Aztec consumer to confirm end-to-end success.

\section{Threats and Limitations}
\label{sec:threats}

\paragraph{Bridge trust and policy.}
Guardian trust and Wormhole finality policies are assumed correct; changes in guardian sets or policy misconfigurations are out of scope.

\paragraph{Parameter binding in current code.}
The legacy \texttt{consumeWithSecret(vaa, secretHash)} lets a relayer choose \texttt{secretHash}.
While this does not break replay/authenticity, an incorrect hash can hinder Aztec consumption until corrected.
The bound variant removes that vector.

\paragraph{Commitment format.}
Domain separation is recommended; it reduces cross-context collision risks.
If user payloads are low-entropy, application-level salting is encouraged.

\paragraph{DoS surfaces.}
Relayers can feed malformed VAAs; verification short-circuits these at bounded cost on EVM.
On Aztec, \nolinkurl{consume_from_inbox} should enforce existence before heavier work.

\section{Related Work}
\label{sec:related}

\textit{Foundational systems.}
We rely on Wormhole VAAs for cross-domain attestation \cite{wormhole_docs}, the Aztec protocol for private L2 execution \cite{aztec_docs}, and Solana’s runtime and developer tooling \cite{solana_docs}. See \cite{anchor_lang_crate}.

\textit{Solana-oriented ZK lines.}
Light Protocol’s zk-compression aims at scalable on-chain state commitments and privacy on Solana itself \cite{light_protocol_whitepaper}.
Arcium targets confidential off-chain compute with Solana integrations \cite{arcium_docs}.
Our bridge is complementary: rather than verifying heavy proofs on Solana, it dispatches private work to Aztec with strong origin binding and replay prevention.

\textit{Coprocessors and zkVMs on EVM.}
Succinct provides general coprocessor patterns and proof verification \cite{succinct_docs}.
RISC Zero exposes a zkVM for arbitrary guest code with end-to-end tooling \cite{risc0_docs}.
Axiom proves statements over Ethereum state with programmable queries \cite{axiom_docs}.
These are broader in compute expressivity; our contribution is a minimal, replay-safe bridge pattern that binds parameters at the message layer and slots into Aztec’s L1$\to$L2 inbox.

\textit{Proof systems.}
PLONK and pairing-based NIZKs are covered in \cite{plonk_eprint_2019}. See also \cite{groth_2016}. For FRI-based IOPPs, see \cite{fri_iopp_2018}. See \cite{deep_fri_2019}.

\section{Conclusion}
\label{sec:conclusion}

We provided a formal specification of a Solana$\to$Aztec ZK coprocessor bridge via Wormhole and an EVM Portal, with explicit state machines, normative encodings, and proof sketches for replay-safety, origin authenticity, finality alignment, parameter binding, privacy, idempotence, and liveness.
We complemented the core with engineering guidance and a concise reproducibility note to replicate a public-testnet run.
This bridges academic clarity and practical deployment for replay-safe private dispatch.

\begingroup
\sloppy\emergencystretch=3em
\Urlmuskip=0mu plus 2mu\relax
\def\UrlBreaks{\do\/\do\-\do\_\do\.\do\?\do\&\do\=\do\#}

\endgroup

\end{document}